\documentclass[aps,preprintnumbers,floatfix,article,amsmath,amssymb,floatfix,10pt,prd,superscriptaddress,nofootinbib]{revtex4-2}
\usepackage{bm}
\usepackage{amsfonts}
\usepackage{latexsym}
\usepackage[latin1]{inputenc}
\usepackage{graphicx}
\usepackage{amsmath}
\usepackage{palatino}
\usepackage{mathpazo}
\usepackage{textcomp}
\usepackage{float}
\usepackage{booktabs}
\usepackage{dcolumn}
\usepackage{ragged2e}
\usepackage{hyperref}
\hypersetup{colorlinks,citecolor=blue}
\hypersetup{colorlinks=true,linkcolor=red,filecolor=magenta,    urlcolor=cyan}
\usepackage{amsmath}
\usepackage{xcolor}
\usepackage{orcidlink}
\usepackage{epsfig}
\usepackage{caption}
\usepackage{subcaption}
\usepackage{commath}
\def\jnl@style{\it}
\def\aaref@jnl#1{{\jnl@style#1}}

\def\aaref@jnl#1{{\jnl@style#1}}

\def\aj{\aaref@jnl{AJ}}                   
\def\apj{\aaref@jnl{ApJ}}                 
\def\apjl{\aaref@jnl{ApJ}}                
\def\apjs{\aaref@jnl{ApJS}}               
\def\apss{\aaref@jnl{Ap\&SS}}             
\def\aap{\aaref@jnl{A\&A}}                
\def\aapr{\aaref@jnl{A\&A~Rev.}}          
\def\aaps{\aaref@jnl{A\&AS}}              
\def\mnras{\aaref@jnl{Mon.~Not.~Roy.~Astron.~Soc.}}             
\def\prd{\aaref@jnl{Phys.~Rev.~D}}        
\def\prc{\aaref@jnl{Phys.~Rev.~C}}  
\def\prl{\aaref@jnl{Phys.~Rev.~Lett.}}    
\def\qjras{\aaref@jnl{QJRAS}}             
\def\skytel{\aaref@jnl{S\&T}}             
\def\ssr{\aaref@jnl{Space~Sci.~Rev.}}     
\def\zap{\aaref@jnl{ZAp}}                 
\def\nat{\aaref@jnl{Nature}}              
\def\aplett{\aaref@jnl{Astrophys.~Lett.}} 
\def\apspr{\aaref@jnl{Astrophys.~Space~Phys.~Res.}} 
\def\physrep{\aaref@jnl{Phys.~Rep.}}      
\def\physscr{\aaref@jnl{Phys.~Scr}}       
\def\commat{\aaref@jnl{Comm.~Math.~Phys.}}              
\def\science{\aaref@jnl{Science}}               
\def\cqg{\aaref@jnl{Classical Quant.~Grav.}}            
\def\jpcs{\aaref@jnl{JPCS}}                                     
\def\ijmpd{\aaref@jnl{Int.~J.~Mod.~Phys.~D}}                    
\def\grg{\aaref@jnl{Gen.~Relat.~Gravit.}}               
\def\rpp{\aaref@jnl{Rep.~Prog.~Phys.}}          
\def\npa{\aaref@jnl{Nucl.~Phys.~A}}        
\def\lrr{\aaref@jnl{Living Rev.~Rel.}}                   
\def\jcap{\aaref@jnl{J.~Cosmology Astropart.~Phys.}}    
\def\rmp{\aaref@jnl{Rev.~Mod.~Phys.}}   
\def\epjc{\aaref@jnl{Eur.~Phys.~J.~C}} 
\def\plb{\aaref@jnl{~Phy.~Lett.~B}} 
\def\mpla{\aaref@jnl{Mod.~Phy.~Lett.~A}} 
\def\arxiv{\aaref@jnl{arxiv.org}}

\allowdisplaybreaks[1]

\begin{document}
\title{\bf Bouncing cosmological models in a functional form of $F(R)$ gravity}

\author{A. S. Agrawal\orcidlink{0000-0003-4976-8769}}
\email{agrawalamar61@gmail.com}
\affiliation{Department of Mathematics,
Birla Institute of Technology and Science-Pilani, Hyderabad Campus,
Hyderabad-500078, India.}

\author{S. Mishra}
\email{sachidanandamishra1998@gmail.com}
\affiliation{Department of Physics, Indira Gandhi Institute of Technology, Sarang, Dhenkanal, Odisha-759146, India.}

\author{S.K. Tripathy\orcidlink{0000-0001-5154-2297}}
\email{tripathy\_sunil@rediffmail.com}
\affiliation{Department of Physics, Indira Gandhi Institute of Technology, Sarang, Dhenkanal, Odisha-759146, India.}

\author{B. Mishra\orcidlink{0000-0001-5527-3565}}
\email{bivu@hyderabad.bits-pilani.ac.in}
\affiliation{Department of Mathematics,
Birla Institute of Technology and Science-Pilani, Hyderabad Campus,
Hyderabad-500078, India.}

\begin{abstract}
\textbf{Abstract}
We have investigated some bouncing cosmological models in an isotropic and homogeneous space time with the $F(R)$ theory of gravity. Two functional forms of $F(R)$ have been investigated with a bouncing scale factor. The dynamical parameters are derived and analysed along with the cosmographic parameters. The analysis in both the models show the occurrence of bouncing scenario. The violation of strong energy conditions in both models is also shown. In the stability point of view we have analysed the behaviour of $F_{R}=\frac{dF}{dR}$ with respect to cosmic time and both the models exhibit stable behaviour.\\
\end{abstract}
\maketitle

\section{Introduction}
The initial singularity is another important issue that General Relativity (GR) has encountered among other issues during early Universe. Friedmann \cite{Friedmann22, Friedmann24} claimed that the occurrence of initial singularity was during the beginning of the evolution of Universe. It is believed that singularity issue occurred before the inflation, because the inflationary scenario resolved certain key issues of early Universe \cite{Brout78, Guth81, Starobinsky80}. One possible solution might be the Universe does not attained singularity during the contraction, but expands after experiencing a bounce. This concept is known as the big bounce. Recent discoveries \cite{Riess98, Perlmutter99, Tegmark04, Abazajian04, Spergel03, Hinshaw13, Parkinson12}, have revealed that our Universe is undergoing a late time accelerated expansion phase, which is explained by dark energy, time-independent vacuum energy (according to the $\Lambda$CDM model). The cosmological constant \cite{Weinberg89}, scalar fields (including quintessence, phantom, quintom, tachyon, and others) \cite{Kamenshchik01, Caldwell02, Amani11a, Sadeghi09, Setare09a, Setare09b}, and holographic models \cite{Amani11b} are possibilities for describing dark energy scenarios. Modified gravity theory has advantages over other models since it avoids expensive numerical computations and is consistent with current data for a late phase accelerating Universe and dark energy. So the models with such theories are being designed to modify the standard nature of GR by replacing the Ricci scalar $R$ in Einstein-Hilbert action with $f(R)$. Several modified theories of gravity have been developed, such as $f(R)$ gravity \cite{Carroll04a, Nojiri03a, Nojiri07a, Nojiri11a, Nojiri14a, Capozziello11a, Bamba10a, Bamba12a, Amendola07a}, $f(G)$ gravity \cite{Nojiri05}, $f(\mathcal{T})$ gravity \cite{Linder10, Myrzakulov11} and $f(R,T)$ gravity \cite{Harko11, Mishra16, Mishra18, Yousaf16, Velten17, Carvalho17, Tretyakov18, Baffou19, Alhamzawi16, Alves16, Abbas17}, Teleparallel gravity \cite{Abedi18, D'Agostino18}, where $\mathcal{T}$ denotes the torsion scalar and $G$ is the Gauss Bonnet invariant term. Some other important work on modified theories of gravity \cite{Capozziello11, Capozziello19, Mishra18a, Mishra18b} are available in the literature. Most recent $f(Q)$ gravity or symmetric teleparallel gravity \cite{Conroy18} and $f(Q,T)$ \cite{Xu19} gravity have been proposed, where $Q$ and $T$ respectively represent the non-metricity and trace of energy momentum tensor. \\

The inflationary scenario has been challenged, and the matter bounce scenario has been presented as a possible alternative to address the initial singularity issue. The Universe goes through an initial matter dominated contraction phase, then a non singular bounce, and finally a causal generation for fluctuation in the bouncing scenario. For this, the bouncing scenario is a typical example, and a null energy condition (NEC) has to be violated to realize a solution in a spatially flat FLRW metric in GR. The matter bounce scenario has gained a lot of attention among the numerous bouncing models proposed because it creates a scale-invariant power spectrum. Additionally, the Universe passes through a matter-dominated epoch at the late time in a matter bounce scenario. Alternative gravity theories like $f(R)$ gravity \cite{Bamba14b,Barragan09b,Barragan10b,Chakraborty18b}, $f(G)$ gravity \cite{Bamba14c,Bamba15}, $f(R,T)$ gravity \cite{Singh18,Tripathy19,Mishra19,Tripathy21,Agrawal21a}, $f(Q,T)$ gravity \cite{Agrawal21b}, $f(\mathcal{T})$ gravity \cite{Cai11},{\bf $f(Q)$ gravity \cite{Agrawal23} and $f(R,G)$ gravity \cite{Lohakare22} } have all successfully studied bouncing cosmologies. The present work is on the bouncing model in a modified theory of gravity, the $f(R)$ theory in an FLRW space -time. To note, $f(R)$ gravity theory is an excellent alternative to the standard gravity model to study the dark energy cosmological models. In the $f(R)$ modified gravity framework, Odintsov and Oikonomou \cite{Odintsov17} have investigated a bouncing cosmology with a Type IV singularity at the bouncing point. Elizalde et al. \cite{Elizalde20} et al. have studied the extended matter bounce scenario in ghost-free $f(R,G)$ gravity which is compatible with the gravitational waves.\\

$f(R)$ gravity is an important and well-known modified gravity theory \cite{Sotiriou10, Felice10}. Hu and Sawicki proposed a late-time cosmological model \cite{Hu07}, the model can explain the late-time acceleration of the Universe, without the need of dark energy. Starobinsky \cite{Starobinsky07} suggested a $f(R)$ gravity model that is in line with cosmological conditions and accords with laboratory experiments and observations of the solar system. However, the exponential gravity model was developed and studied by Cognola et al. \cite{Cognola08}. This model accurately captures the natural inflation of the early Universe and the accelerated expansion of the present Universe.  Further  these models are studied to get the traversable wormhole solutions gravity by Shamir and Fayyaz \cite{Shamir20}. Chen et al. \cite{Chen19} studied the matter power spectra with the dynamical background evolution in $f(R)$ theory. 

In the context of $f(R)$ gravity, Odintsov et al. \cite{Odintsov20} have proposed a cosmological model that merges a non-singular bounce to a matter-dominated epoch and space-time dominated to a late time accelerating epoch; i.e., the model is similar to a generalized matter bounce model which is also compatible with the late DE dominant phase of the cosmic evolution. Odintsov et al. \cite{Odintsov21} investigated a Chern-Simons corrected $f(R)$ gravity theory of a non-singular bounce to a dark energy epoch, where the Chern-Simons coupling function is supposed to have a power law behaviour with the Ricci scalar. In a gravitational model with curvature-squared $R^{2}$ and curvature quartic $R^{4}$ non-linearities, Saidov and Zhuk \cite{Saidov10} have examined bouncing inflation. Barragan et al. \cite{Barragan09}, have analysed the criteria that guarantee the existence of homogeneous and isotropic models that avoids the Big Bang singularity in Palatini formalism. The modified Friedmann equation in LQC has been transferred to the Palatini $f(R)$ theory \cite{Olmo09} by Olmo and Singh whereas Olmo \cite{Olmo11}, has discovered the necessary $f(R)$ function that must be taken into account to create a bouncing cosmology of this type of LQC. In the generalized $f(R)$ theory, Nojiri et al. \cite{Nojiri19} have studied non-singular bounce cosmology in the context of Lagrange multiplier. As a result, it is discovered that the weak energy and null energy conditions are violates close to the bouncing point. A common $f(R)$ gravity model is used to describe the phenomenology of the current non-singular bounce.
 
In this paper, our objective is to study some bouncing cosmological models to avoid the initial singularity issue with some of the functional forms of $F(R)=R+f(R)$, $f(R)$ is the deviation of $F(R)$ from the Einstein gravity. To explain the late-time cosmic speed-up issue, the models will look at geometrical degrees of freedom. The explanation of $F(R)$ gravity and the derivation of $F(R)$ field equations are presented in Sec. II of the study. In Sec. III, the bouncing scale factor and Hubble parameter were introduced. Two models with the bouncing scale factor and functional form of $F(R)$ are provided in Sec. IV. The cosmographic parameters are discussed in Sec. V and the energy conditions of both models are given in Sec. VI. Stability analysis has been done in Sec. VII. The model results and conclusions are presented in Sec.VIII. \\

\section{Field equations of \texorpdfstring{$F(R)$}{} gravity}
The action for $F(R)$ gravity can be defined as,
\begin{equation}\label{eq.1}
S=\int\sqrt{-g}\frac{F(R)}{2\kappa^{2}}d^{4}x,    
\end{equation}

$\kappa^{2}=\frac{8\pi G}{c^4}$, $G$ be the Newton's gravitational constant, $g$ is determinant of the metric tensor $g_{ij}$. Varying action  \eqref{eq.1} with respect to $g_{ij}$, the $F(R)$ gravity field equations can be obtained as,

\begin{equation}\label{eq.2}
F_{R}R_{ij}-\frac{1}{2}F g_{ij}-\nabla_{i}\nabla_{j}F_{R}+g_{ij}\square F_{R}=0
\end{equation}

Here $F_{R}=\frac{dF}{dR}$, $\nabla_{i}$ represents the covariant derivative, $\square\equiv g^{ij}\nabla_{i}\nabla_{j}$ is the d'Alembert operator. The natural system of unit $8\pi G=\bar{h}=c=1$ has been used, where $G$, $\bar{h}$ and $c$ respectively denote the Newtonian gravitational constant, reduced Planck constant and velocity of light in vacuum respectively. 

We consider the flat FLRW space-time as,
\begin{equation}\label{eq.3}
ds^{2}=-dt^{2}+a^{2}(t)(dx^{2}+dy^{2}+dz^{2}),    
\end{equation}
For this metric, the temporal and spatial components of the Eq. \eqref{eq.2} becomes
\begin{eqnarray}\label{eq.4}
0&=&-\frac{F}{2}+3\left(H^{2}+\dot{H}\right)F_{R}-18\left(4H^{2}\dot{H}+H\ddot{H}\right)F_{RR}\\
0&=&\frac{F}{2}-3\left(H^{2}+\dot{H}\right)F_{R}+6\left(8H^{2}\dot{H}+4\dot{H}^{2}+6H\ddot{H}+\dot{\ddot{H}}\right)F_{RR}+36\left(4H\dot{H}+\ddot{H} \right)^{2}F_{RRR}\label{eq.5}
\end{eqnarray}

where $F_{RR}=\frac{d^{2}F}{dR^{2}}, ~F_{RRR}=\frac{d^{3}F}{dR^{3}}$ and $H=\frac{\dot{a}}{a}$ is the Hubble parameter. When the above equations are compared to the standard Friedmann equations, it is clear that $F(R)$ gravity contributes to the energy-momentum tensor, with its effective energy density $\rho_{eff}$ and pressure $p_{eff}$ given by

\begin{eqnarray}\label{eq.6}
\rho_{\text{eff}}&=&-\frac{f}{2}+3\left(H^{2}+\dot{H}\right)f_{R}-18\left(4H^{2}\dot{H}+H\ddot{H}\right)f_{RR}\\
p_{\text{eff}}&=&\frac{f}{2}-3\left(H^{2}+\dot{H}\right)f_{R}+6\left(8H^{2}\dot{H}+4\dot{H}^{2}+6H\ddot{H}+\dot{\ddot{H}}\right)f_{RR}+36\left(4H\dot{H}+\ddot{H} \right)^{2}f_{RRR}\label{eq.7}
\end{eqnarray}

Eqns. \eqref{eq.6} and \eqref{eq.7} can be expressed in terms of Hubble parameter, $H=\frac{\dot{a}}{a}$ and the derivatives of functional form of $F(R)$ with respect to $R$ in which the Ricci scalar, $R=6\left(\frac{\ddot{a}}{a}+\frac{\dot{a}^{2}}{a^{2}} \right)$. So, we need a Hubble function to obtain the energy density and pressure of the matter field to further study the dynamics of the Universe. Also to study the issue of late time acceleration issue, the equation of state (EoS) parameter behaviour to be analysed, which can be obtained as,
\begin{equation}\label{eq.8}
\omega_{\text{eff}}=\frac{p_{\text{eff}}}{\rho_{\text{eff}}} =-1+\frac{12\left(2H^{2}\dot{H}+4\dot{H}^{2}+3H\ddot{H}+\dot{\ddot{H}}\right)f_{RR}+72\left(4H\dot{H}+\ddot{H} \right)^{2}f_{RRR}}{f-6\left(H^{2}+\dot{H}\right)f_{R}+36\left(4H^{2}\dot{H}+H\ddot{H}\right)f_{RR}}
\end{equation}

Where $f(R)$ represents the departure of $F(R)$ gravity from Einstein gravity, $F(R)=R+f(R)$. As expected, the effective energy-momentum tensor relies on the nature of $F(R)$. So, in the subsequent sections, we will study the bouncing scenario and late time cosmic acceleration issue of the Universe by considering the bouncing scale factor and some of the functional forms of $F(R)$.  

\section{The Scale Factor}
Inflationary cosmology is one of two extant theories of the early Universe, with the other being bounce cosmology, in which the theoretical contradictions of the Big Bang description of our Universe are addressed. The most recent observational data imposed strict limits on inflationary models, confirming the validity of some while ruling out others. Here we intend to study the bouncing scenario in the $F(R)$ theory of gravity.
\begin{itemize}
\item For the case of non-singular bounce, bouncing scenario behaves as a contracting nature formulated by the scale factor which decreases with time, i.e., $\dot{a}<0$, means Hubble parameter is negative in contacting phase, i.e., $H=\dot{a}/a<0$.
\item For bouncing epoch, contracting nature of scale factor to a non zero finite critical size is obtained as a result of which the Hubble parameter vanishes at bounce making $H=0$.
\item Nature of scale factor increases with time in a positive acceleration, so as the Hubble parameter becomes positive after the bounce, i.e., $\dot{a}>0$.
\item For the situation of near to bouncing epoch, Hubble parameter holds true, i.e., $\dot{H}>0$  which is suitable for ghost (phantom) behaviour of the model.
\item Also to appreciate a bouncing model, EoS evolves at such phantom region and changes twice, one before the bounce and another after the bounce.
\end{itemize}

So, here our bouncing model with an assumed scale factor obeys above bouncing conditions and the simultaneous dynamical behaviour. Hence we consider a bouncing scale factor $a(t)=\left(\frac{\alpha}{\beta}+t^{2}\right)^{\frac{1}{2\beta}}$, where $\alpha$ and $\beta$ are positive constants and subsequently, the Hubble parameter $H=\frac{t}{\alpha +\beta t^{2}}$. 
\begin{figure} [H]
\minipage{0.45\textwidth}
\includegraphics[width=\textwidth]{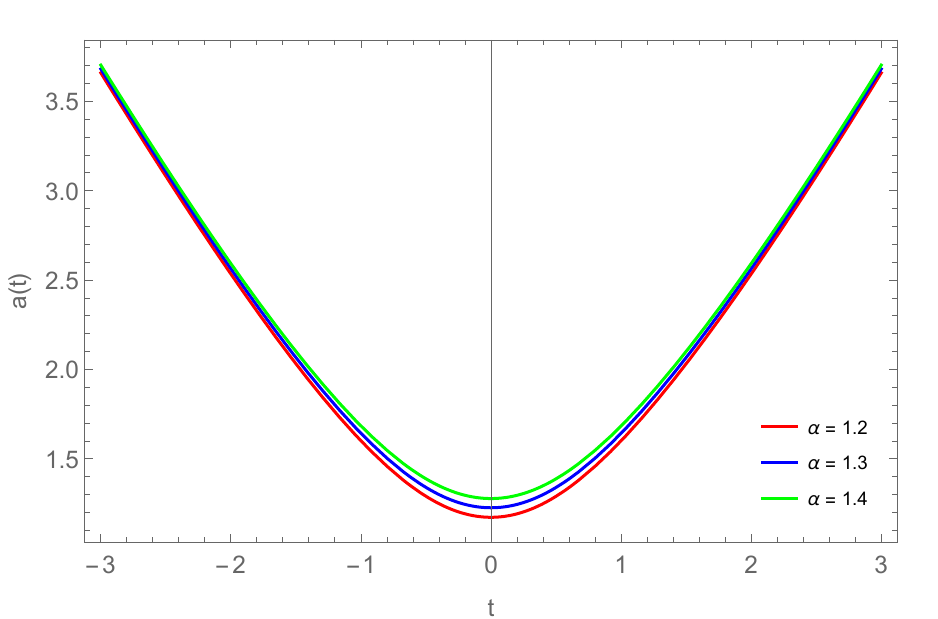}
\endminipage\hfill
\minipage{0.45\textwidth}
\includegraphics[width=\textwidth]{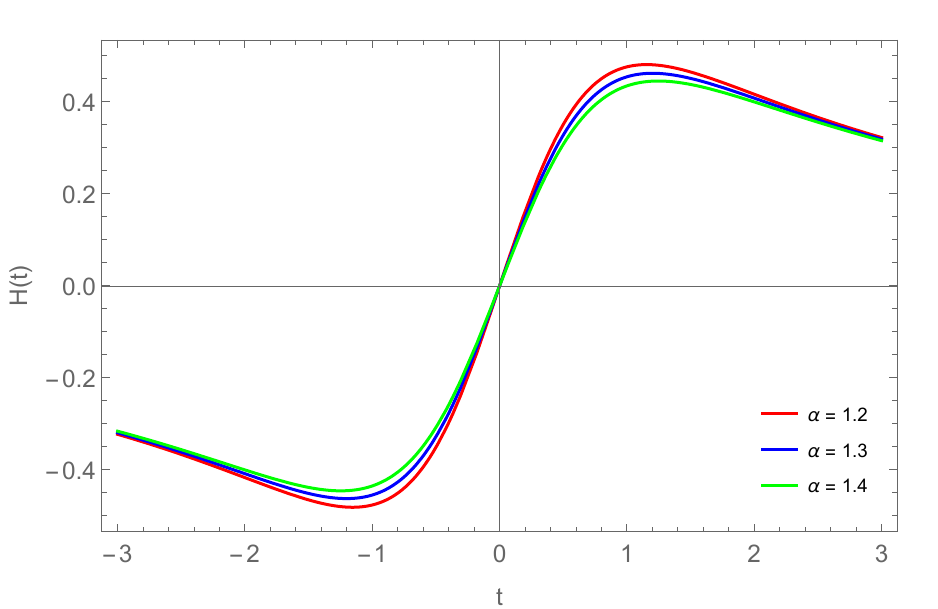}
\endminipage\hfill

\caption{Variation of scale factor  (left panel), Hubble parameter (right panel) in cosmic time with varying $\alpha$ and $\beta = 0.9$.}
\label{Fig.1}
\end{figure}

FIG. \ref{Fig.1} represents the behaviour of scale factor and Hubble parameter with the representative values of the parameter $\alpha=1.2, 1.3, 1.4$. It has been observed that the bounce occurs at $t=0$ and the parameter $\alpha$ controls the slope of the curve. A higher value of $\alpha$ yields a higher slope. The bounce appears to be symmetric, the scale factor appears to decrease from a higher value at early time ( in the negative time domain) bounces at $t=0$ and increases further at late time. The curve of Hubble parameter increases from a higher negative value crosses the bouncing point at $t=0$ and increases further over the evolution. The behaviour of the parameters support the occurrence of bouncing trajectory, thereby to avoid the initial singularity issue. 

While considering bouncing scenarios within the purview of $F(R)$ gravity theory, it is useful to consider the generation era of the perturbation modes. Usually for many bouncing models considered through the choice of some of the scale factors, the Hubble parameter vanishes at the bounce epoch which leads to the divergence of the comoving Hubble radius, defined by $r_h=1/aH$.  The asymptotic behaviour of the comoving Hubble radius, on the other hand, shows the accelerating or decelerating nature of the Universe. For some specific choices of the scale factors, the Hubble radius drops monotonically on both sides of the bounce before asymptotically shrinking to zero. Such a behaviour indicates an accelerating  Universe at late times. As a result, in such instances, the Hubble horizon shrinks to zero for large values of cosmic time, and only the Hubble horizon has an infinite size near the bouncing point. However, for some other choices of the bouncing scale factors, the Hubble radius diverges at late time indicating a decelerating Universe. In such scenarios, the perturbation modes are created at very large negative cosmic times, corresponding to the low curvature regime of the contracting era, rather than near the bouncing era as in prior cases. As a result, the primordial perturbation modes relevant to the present time era are formed for cosmic times near the bouncing point, because all the primordial modes are contained in the horizon only at that time. The modes escape the horizon when the horizon shrinks and become relevant for present-day observations \cite{Odintsov20a}. where the Hubble radius approaches zero asymptotically because all of the perturbation modes are within the horizon at that time, the perturbations occur near the bounce. The Planck restrictions \cite{Ade14,Ade16,Akrami20} become compatible with the $F(R)$ gravity theory. Odintsov et al., \cite{Odintsov20b} discovered that $F(R)$ gravity leads to feasible bounce only when perturbations generate near the bounce directly from observational indices using a bottom-up approach. In order to check whether, the scale factor considered in the present work is in conformity with the generation of perturbation modes and whether it is viable to consider such a bouncing scenario within the framework of $F(R)$ gravity, we have shown the cosmic Hubble radius as a function of time for different choices of the parameter $\alpha$ and a specific choice of $\beta=0.9$. With these choices, we obtain that, the cosmic Hubble radius monotonically decreases symmetrically around the bouncing epoch and tends to zero asymptotically in both the positive and negative time domain. This finding is in conformation with that of Odintsov  et al. in Ref. \cite{Odintsov20a}, where they have mentioned that the cosmic Hubble radius for the choice of the scale factor $a_F(t_F)=\left(a_0t_F^2+1\right)^{n}$ drops monotonically on both sides of the bounce for $n>1/2$. In view of this, we may infer that, our present bouncing model may be compatible with Planck constraints.

In the following sections, we will consider two different functional forms of $F(R)$  and the given symmetric bouncing scale factor to obtain two different bouncing scenarios.

\begin{figure}
\centering
\includegraphics[width=0.5\textwidth]{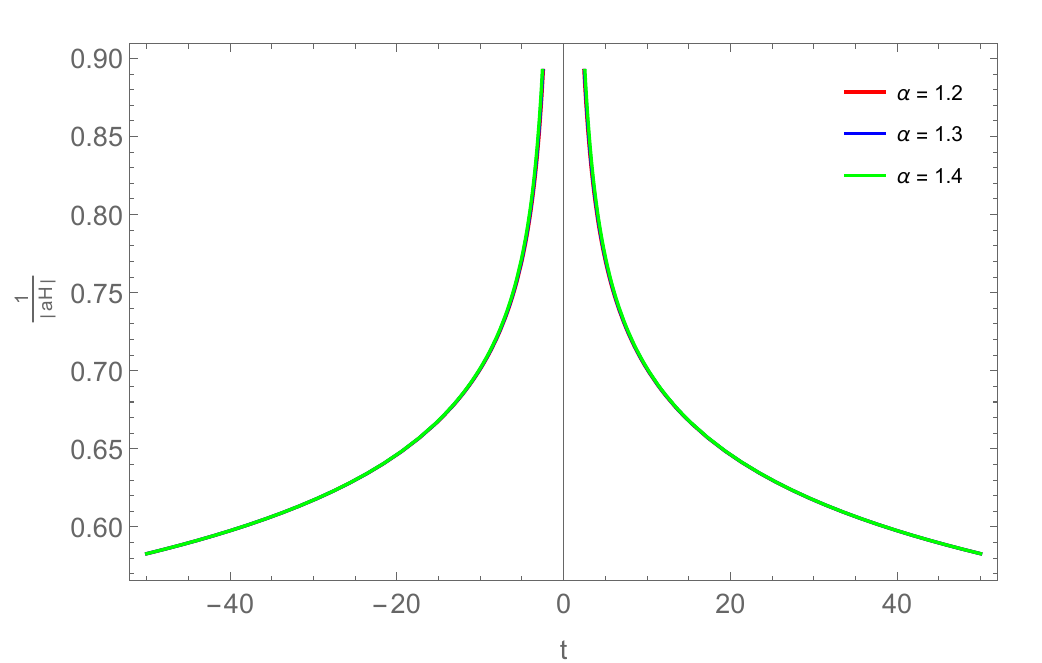}
\caption{Variation of the Hubble radius in cosmic time with varying $\alpha$ and $\beta = 0.9$.}
\label{Fig.1a}
\end{figure}
\section{Models}
We need to have a functional form of $F(R)$ here, thus we will use two well-known functional forms as Model I and Model II,

\subsection{Model I}
To construct the cosmological model, Eq. \eqref{eq.6} and Eq. \eqref{eq.7} are required to be solved, so that the dynamical parameters can be obtained. To do so, a functional form for $F(R)$ to be considered. We consider the form of $F(R)$ \cite{Starobinsky07,Chen19} as,
\begin{eqnarray}\label{eq.9}
F(R)=R+\lambda R_{0}\left[\left(1+\frac{R^{2}}{R_{0}^{2}}\right)^{-n}-1\right],  \end{eqnarray}
where, $R_{0}$ is the constant characteristic curvature, $\lambda$ and $n$ are also constants. We chose the value of the exponent, $n=1$ in order to match with the Starobinsky model.  Using the functional form of $F(R)$ \eqref{eq.9}, Eqs. \eqref{eq.6}, \eqref{eq.7} and \eqref{eq.8} can be reduced respectively as,
\begin{eqnarray}
\rho_{\text{eff}}&=& \frac{\lambda  R_{0} \left(72 H R_{0}^2 \left(R_{0}^2-3 R^2\right) \left(4 H \dot{H}+\ddot{H}\right)-12 R_{0}^2 R \left(\dot{H}+H^2\right) \left(R_{0}^2+R^2\right)+R^2 \left(R_{0}^2+R^2\right)^2\right)}{2 \left(R_{0}^2+R^2\right)^3}\label{eq.10}\\
p_{\text{eff}}&=&\frac{4 \lambda  R_{0}^3 R \left(\dot{H}+3 H^2\right) \left(R_{0}^2+R^2\right)^{2}-24 \lambda  R_{0}^3 \left(R_{0}^2-3 R^2\right) \left(R_{0}^2+R^2\right) \left(4 \dot{H} \left(\dot{H}+2 H^2\right)+6 H \ddot{H}+\dot{\ddot{H}}\right)}{2 \left(R_{0}^2+R^2\right)^4}\nonumber \\ &&+\frac{1728 \lambda  R_{0}^3 R (R_{0}-R) (R_{0}+R) \left(4 H \dot{H}+\ddot{H}\right)^2-\lambda  R_{0} R^2 \left(R_{0}^2+R^2\right)^3}{2 \left(R_{0}^2+R^2\right)^4}\label{eq.11}\\
\omega_{\text{eff}}&=&\frac{4 R_{0}^2 R \left(\dot{H}+3 H^2\right) \left(R_{0}^2+R^2\right)^2+1728 R_{0}^2 R (R_{0}-R) (R_{0}+R) \left(4 H \dot{H}+\ddot{H}\right)^2-R^2 \left(R_{0}^2+R^2\right)^3}{\left(R_{0}^2+R^2\right) \left(72 H R_{0}^2 \left(R_{0}^2-3 R^2\right) \left(4 H \dot{H}+\ddot{H}\right)-12 R_{0}^2 R \left(\dot{H}+H^2\right) \left(R_{0}^2+R^2\right)+R^2 \left(R_{0}^2+R^2\right)^2\right)}\nonumber \\&&-\frac{24 R_{0}^2 \left(R_{0}^2-3 R^2\right) \left(R_{0}^2+R^2\right) \left(4 \dot{H} \left(\dot{H}+2 H^2\right)+6 H \ddot{H}+\dot{\ddot{H}}\right)}{\left(R_{0}^2+R^2\right) \left(72 H R_{0}^2 \left(R_{0}^2-3 R^2\right) \left(4 H \dot{H}+\ddot{H}\right)-12 R_{0}^2 R \left(\dot{H}+H^2\right) \left(R_{0}^2+R^2\right)+R^2 \left(R_{0}^2+R^2\right)^2\right)}\label{eq.12}
\end{eqnarray}

\begin{figure}[H]
\centering
\minipage{1\textwidth}
\includegraphics[width=\textwidth]{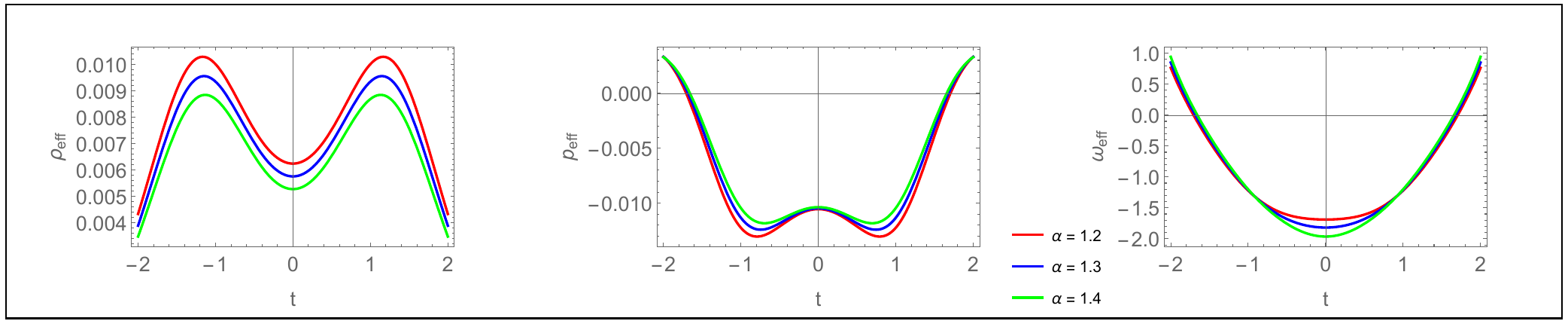}
\endminipage\hfill
\caption{Variation of effective energy density (left panel), effective pressure (middle panel), EoS parameter (right panel) in cosmic time with varying $\alpha$ and $\beta =0.9$, $R_{0}=2$, $\lambda =0.01$, $n=1$ for Model I.}
\label{Fig.3}
\end{figure}
For Model I, the effective pressure remains negative near the  bouncing point i.e., $t=0$ of the evolution. The energy density remains in the positive region throughout the evolution. For higher values of $\alpha$, the bounce at $t=0$ becomes more prominent. The energy density increasing initially, shows a kind of ditch reduces near and at the bounce and then subsequently decreases. The EoS parameter decreases in the contracting phase, crosses phantom-divide line. The EoS parameter shows the phantom like behaviour at bounce epoch and it gets increased in the expanding phase of the evolution. It is also showing the symmetric behaviour and crosses the phantom-divide two times, once in the pre bounce and other in the post bounce phase. It is evident that the current model mostly in the quintessence area, while exhibiting phantom-like behaviour at bounce epoch, and $\Lambda$CDM behaviour in both positive and negative time scales as we move away from bounce epoch. The size of the well occurred here depends on the value of $\alpha$, more the value of $\alpha$ deeper is the well. When the value of $\alpha$ is small, only near the bounce the well is visible, else it remains mostly in the quintessence phase.

\subsection{Model II}
We consider another form of the function $F(R)$ \cite{Cognola08} as, 
\begin{eqnarray}\label{eq.13}
F(R)=R+R_{0}\lambda\left(e^{-\frac{R}{R_{0}}}-1\right),
\end{eqnarray}
where $R_0$ and $\lambda$ are constants. The same scale factor has been considered here as in Model I. From Eqs. \eqref{eq.6} and \eqref{eq.7}, the effective energy density, effective pressure and EoS parameter for the exponential $F(R)$ form \eqref{eq.13} can be obtained as,
\begin{eqnarray}
\rho_{\text{eff}}&=&\frac{\lambda  \left(e^{-\frac{R}{R_{0}}} \left(-6 \left(24 H^2+R_{0}\right) \dot{H}-R_{0} \left(6 H^2+R_{0}\right)-36 H \ddot{H}\right)+R_{0}^2\right)}{2 R_{0}}\label{eq.14}\\
p_{\text{eff}}&=&\frac{\lambda  e^{-\frac{R}{R_{0}}} \left(R_{0} \left(R_{0} \left(6 H^2-R_{0} e^{R/R_{0}}+R_{0}\right)+12 \dot{\ddot{H}}\right)+2 \dot{H} \left(R_{0} \left(48 H^2+R_{0}\right)-288 H \ddot{H}\right)+48\dot{H}^2 \left(R_{0}-24 H^2\right)\right)}{2 R_{0}^2}\nonumber 
\\ &&+\frac{72\lambda  e^{-\frac{R}{R_{0}}}\left( H\ddot{H}R_{0}- \ddot{H}^2\right)}{2 R_{0}^2}\label{eq.15}\\
\omega_{\text{eff}}&=&-1-\frac{4 \left(\dot{H} \left(R_{0} \left(12 H^2+R_{0}\right)+144 H \ddot{H}\right)+12\dot{H}^2 \left(24 H^2-R_{0}\right)-3 \left(3 H R_{0} \ddot{H}+\dot{\ddot{H}} R_{0}-6\ddot{H}^2\right)\right)}{R_{0}^3 e^{R/R_{0}}-R_{0} \left(6 \left(24 H^2+R_{0}\right) \dot{H}+R_{0} \left(6 H^2+R_{0}\right)+36 H \ddot{H}\right)}\label{eq.16}
\end{eqnarray}

\begin{figure}[H]
\centering
\minipage{1\textwidth}
\includegraphics[width=\textwidth]{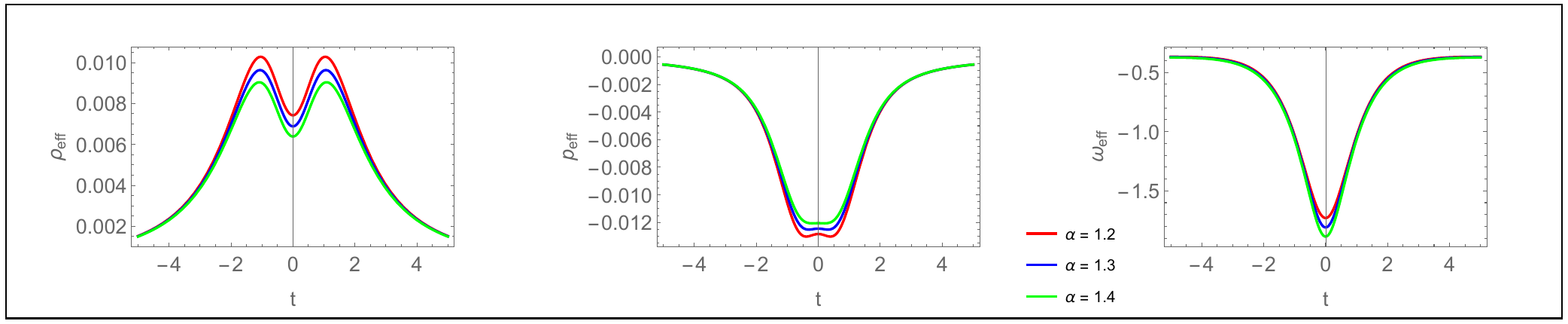}
\endminipage\hfill
\caption{Variation of effective energy density (left panel), effective pressure (middle panel), and EoS parameter (right panel) in cosmic time with varying $\alpha$, $\beta =0.9$, $R_{0}=2.5$, $\lambda=0.01$ for Model II.}
\label{Fig.4}
\end{figure}
In Fig. \ref{Fig.4}, we observe the behaviour of the effective pressure remains entirely negative throughout the evolution of the Universe. The energy density remains entirely positive, and a well appears at the bounce point for a lower $\alpha$. The EoS parameter mostly remains in the phantom phase near the bounce epoch. It crosses the phantom phase as required in bounce model twice, before and after the bounce. As in Model I, the EoS parameter gets a deeper well for higher values of $\alpha$.  It has phantom-like behaviour at bounce epoch passing through phantom divide then showing quintessence behaviour as it moves away from bounce epoch. 

\section{Cosmographic Parameters}
\begin{figure}[H]
\centering
\minipage{0.45\textwidth}
\includegraphics[width=\textwidth]{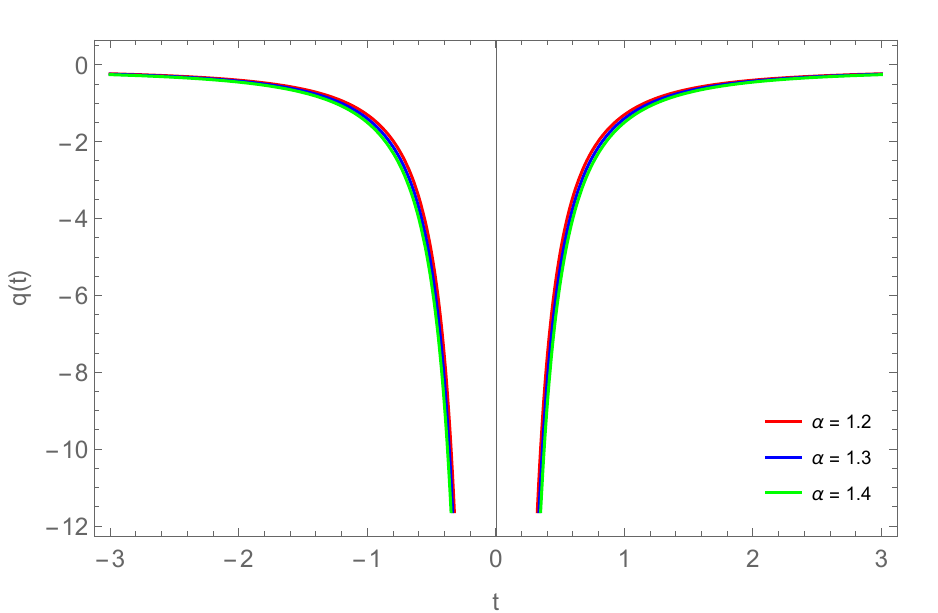}
\endminipage\hfill
\minipage{0.45\textwidth}
\includegraphics[width=\textwidth]{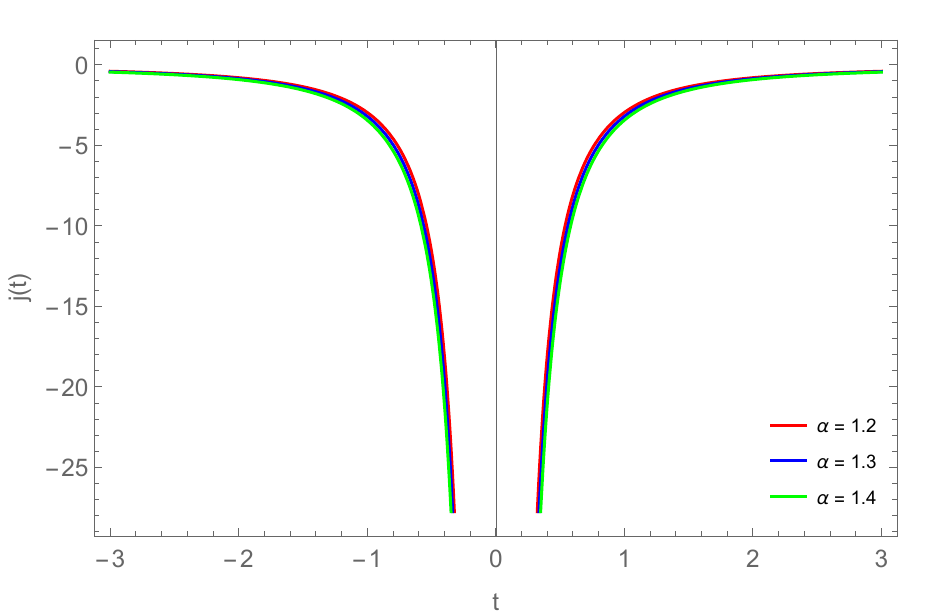}
\endminipage\hfill
\minipage{0.45\textwidth}
\includegraphics[width=\textwidth]{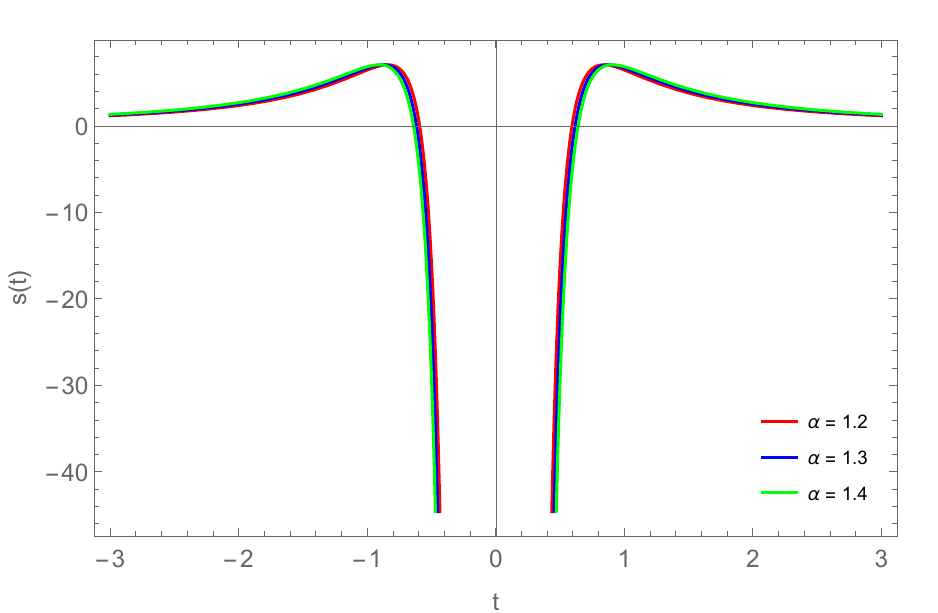}
\endminipage\hfill
\minipage{0.45\textwidth}
\includegraphics[width=\textwidth]{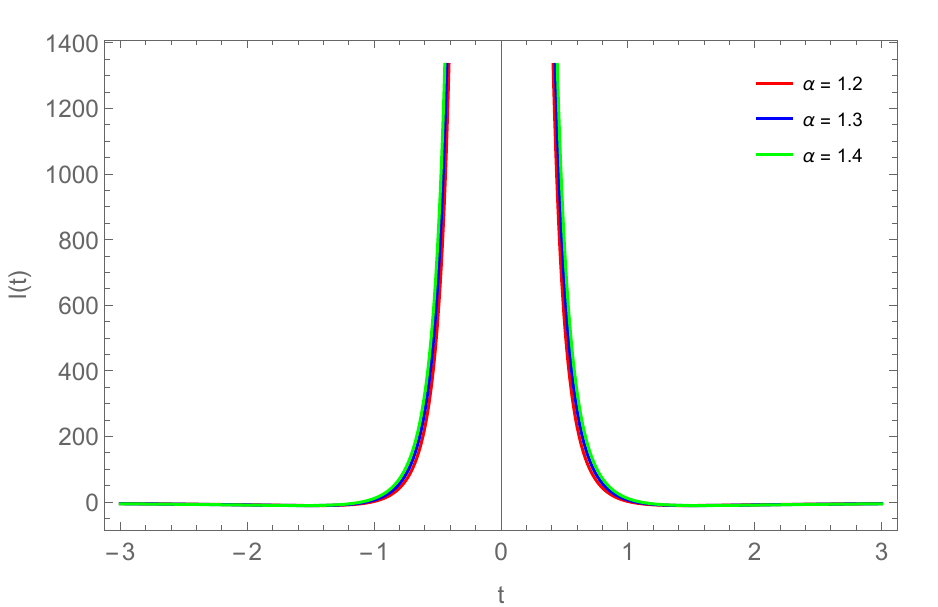}
\endminipage\hfill
\caption{Variation of deceleration parameter (q), jerk parameter (j), snap parameter (s), lerk parameter (l) in cosmic time with varying $\alpha$, $\beta=0.9$}
\label{Fig.5}
\end{figure}

In cosmology research, two families of models are being studied, the dark energy models and modified gravity models. Both are fundamentally different in the sense that, it is possible to distinguish between these models of two families those are having same cosmic expansion history. In an usual manner, the  growth rate of cosmological density perturbations are calculated, and even if the models have identical expansion history, it distinguishes the models depending on different gravity theory. One approach in discriminating dark energy and modified gravity models is with the use of growth factor of matter density perturbation \cite{Linder05}. Another approach of distinguishing the dark energy models is with the state finder pair $(j,s)$ \cite{Sahni03}. It is known that, the expansion rate of the Universe, can be expressed with respect to the scale factor and the deceleration parameter ($q$) corresponds to the second derivative of the scale factor. The jerk parameter ($j$) and snap parameter ($s$) correspond to the third and fourth derivative of the scale factor whereas the fifth derivative is associated with the lerk parameter ($l$). These quantities can be well defined in the Taylor series expansion around the scale factor as,
\begin{equation}\label{eq.20}
a(t)=a(t_{0})+{\sum_{n=0}^{\infty}\frac{1}{n!}\frac{d^{n}a}{dt^{n}}~ \vline}_{~t=t_{0}}(t-t_{0})^{n},
\end{equation}
where $t_0$ is the present cosmic time and $n = 1, 2, 3, . . .$, is an integer. The coefficients of the expansion will give these parameters, called the cosmographic parameters. We can  derive these geometrical parameters from the scale factor as,  
\begin{eqnarray}
q&=&-\frac{\ddot{a}}{a}.\frac{1}{H^2}=\frac{\alpha}{t^{2}}+\beta, \nonumber \\
j&=&\frac{\dddot{a}}{a}.\frac{1}{H^3}=\frac{(2 \beta -1) \left[t^2 (\beta -1)-3 \alpha \right]}{t^2}, \nonumber \\
s&=&\frac{a^{(4)}}{a}.\frac{1}{H^4}=-\frac{(2 \beta -1) \left[3 \alpha ^2+t^4 (\beta -1) (3 \beta -1)+6 \alpha  t^2 (1-3 \beta )\right]}{t^4}, \nonumber\\
l&=&\frac{a^{(5)}}{a}.\frac{1}{H^5}= \frac{\left(8 \beta ^2-6 \beta +1\right) \left[15 \alpha ^2+t^4 (\beta -1) (3 \beta -1)+10 \alpha  t^2 (1-3 \beta )\right]}{t^4}. \label{eq.21}
\end{eqnarray}
The graphical behaviour of the cosmographic parameters are presented in FIG. \ref{Fig.5}. All these parameters are symmetric around the bounce point and experience singularity at the bounce epoch irrespective of the representative values of $\alpha$. The positive value of the deceleration parameter indicates a decelerated Universe, whereas the negative value indicates an accelerated Universe. The deceleration parameter at early and late time, approaches to $-0.6$, thereby confirms the accelerating behaviour of the models and aligned with the present observational value of $q$. According to the graphical behaviour at late time (if t is cosmic time in Gyrs, the value of the deceleration parameter tends to $-0.6$ at $13.8$ Gyrs), which matches with the current observational value of the deceleration parameter at present age of the Universe \cite{Capozziello20}.  It entirely remains in the negative domain, initial decreases and after experiencing the singularity at the bounce increases subsequently, and settled at $-0.6$. The jerk parameter exhibits negative behaviour throughout and evolves from a large value decreases rapidly to experience singularity and again increases drastically. At the same time, the snap parameter exhibits singular behaviour in the negative profile at the bounce epoch; when we move away from the bounce epoch, it crosses the null value and merges at zero again in between reaching its maximum value. The lerk parameter shows opposite behaviour that of snap parameter.

\section{Energy conditions}
In GR, Einstein's field equations address the causal metric and geodesic structure of the space-time, so the energy momentum tensor has to satisfy some conditions. For a space-time $(-,+,+,+)$, we can take the time like vector $u^{i}$ to be normalized as $u_{i}u^{i}=-1$ and the future directed null $k^{i}$ as $k^{i}k_{i}=0$. We can define the energy conditions as the contractions of time like or null vector fields with respect to Einstein tensor and the energy momentum tensor from the matter side of Einstein's field equations \cite{Hawking27,Raychaduhary55, Capozziello18}. We can obtain four energy conditions: 
\begin{itemize}
\item At each point of the space time, the energy momentum tensor should satisfy, $T_{ij}u^{i}u^{j}\geq 0$: Weak Energy Condition (WEC). So, $\rho_{\text{eff}}\geq 0$, $\rho_{\text{eff}}+p_{\text{eff}}\geq 0$.
\item   For the future directed null vector $k^{i}$, $T_{ij}u^{i}u^{j}\geq 0$: Null Energy Condition (NEC). So, $\rho_{\text{eff}}+p_{\text{eff}}\geq 0$. 
\item  The matter flows along time like or null line and with contracted energy momentum tensor, the quantity $-T_{i}^{j} u^{ij}$ becomes future directed time like or null like vector field: Dominant  Energy Condition (DEC). So, $\rho_{\text{eff}}-p_{\text{eff}}\geq 0$.
\item $\left(T_{ij}-\frac{1}{2}T g_{ij}\right)u^{i}u^{j}\geq 0$ says the gravity has to be attractive: Strong Energy Condition (SEC). So, $\rho_{\text{eff}}+3p_{\text{eff}}\geq 0$.
\end{itemize}
The extended theories of gravity are the straight forward extension of Einstein's GR, and so the $F(R)$ gravity. Any such extended theory should be confronted with the energy conditions.

\subsection{Model I}

The energy conditions of the bouncing $F(R)$ model can be obtained by using \eqref{eq.10} and \eqref{eq.11} 

\begin{figure}[H]
\centering
\minipage{1\textwidth}
\includegraphics[width=\textwidth]{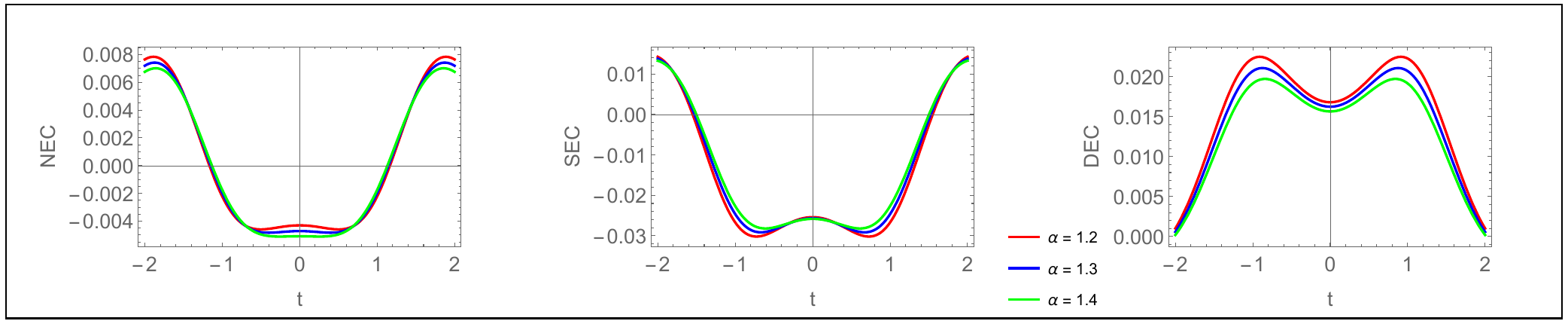}
\endminipage\hfill
\caption{Variation of null energy condition (left panel), strong energy condition (middle panel) and dominant energy condition (right panel) in cosmic time with varying $\alpha$ with $\beta=0.9$, $R_{0}= 2$, $\lambda =0.01$}
\label{Fig.6}
\end{figure}
Graphically, the energy conditions are represented in FIG. \ref{Fig.6} with varying $\alpha$. At the bounce, both the NEC and SEC are violated and DEC is satisfied. The symmetric behaviour around the bounce has been obtained in all energy conditions.  The NEC shows a transition behaviour, mostly it remains in the positive domain both in negative and positive time zone, but near the bounce epoch it remains in the negative domain. Violation of NEC at the bounce, realizes the bouncing model \cite{Nojiri19}. The SEC is violated at bounce and near the bounce epoch during the evolution. The violation of SEC is another requirement for the extended theory of gravity, hence we claim that the model under discussion also favour the late time cosmic acceleration. As expected the DEC satisfies entirely, it increases in the negative time zone and decreases post bounce. In addition, the energy conditions are in good accordance with the EoS parameter in the sense that NEC violates for $\omega_{\text{eff}} \leq -1$, SEC violates for $\omega_{\text{eff}} \leq -1/3$ and DEC satisfies, ${\omega}_{\text{eff}}\leq 1$. This enable us to further claim the validity of the model in the context of  recent cosmic dynamics.

\subsection{Model II}
The energy conditions of Model II can be obtained from \eqref{eq.14} and \eqref{eq.15}

\begin{figure}[H]
\centering
\minipage{1\textwidth}
\includegraphics[width=\textwidth]{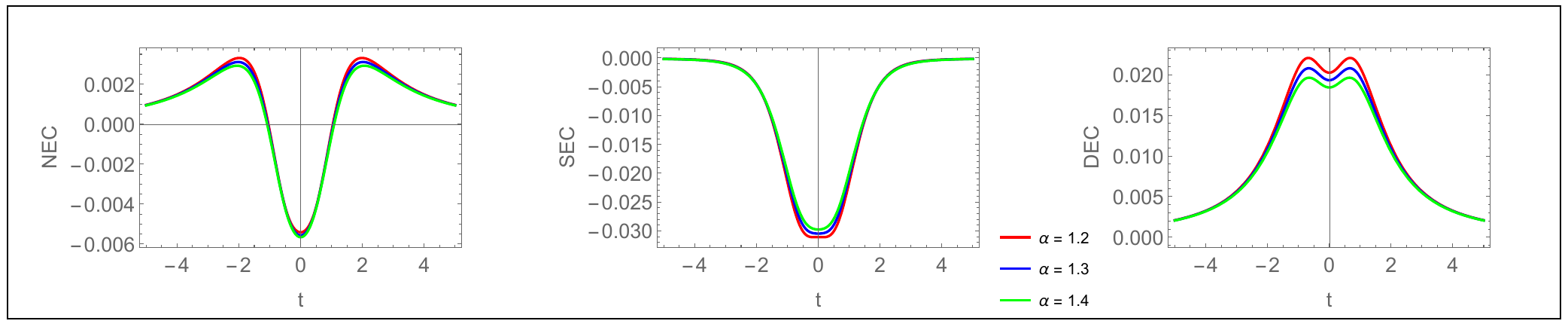}
\endminipage\hfill
\caption{Variation of null energy condition (left panel), strong energy condition (middle panel) and dominant energy condition (right panel) in cosmic time with varying $\alpha$ with $\beta=0.9$, $R_{0}= 2.5$, $\lambda =0.01$}
\label{Fig.7}
\end{figure}
FIG. \ref{Fig.7} depicts the behaviour of energy conditions of Model II. The model violates  NEC at the bounce epoch and while moving away from the bouncing epoch, it fails to violate both in negative and positive time zone, a similar kind of behaviour has been achieved by Nojiri et al., \cite{Nojiri19}. The SEC is violated entirely, it decreases initially and after the bounce increases. In contrast, the DEC fails to violate throughout the evolution. This result of the energy conditions, can confirm the bouncing behaviour of the model and its validity in an extended gravity. It has been observed that the violation of energy conditions depend on the parametric value $\alpha$. For the smaller value of $\alpha$, the well is more deeper at bounce epoch, while for the larger value of $\alpha$, the bounce looks like flat.
\section{Stability Analysis}
The model parameters $(R_{0},\lambda)$ are considered as $(2,0.01)$ and $(2.5,0.01)$ for Starobinsky and exponential gravity model respectively and the scale factor parameters $\beta=0.9$ and $\alpha$ having three value respectively $1.2,1.3$ and $1.4$.  We are now analysing the behaviour of $F_{R}$ with respect to cosmic time to assess the stability of the specified model for the given sets of model parameters. It is noted that the $F_{R}>0$ near the bounce epoch for both models, implying that the models exhibit stable behaviour. \cite{Odintsov20b}
\begin{figure}[H]
\centering
\minipage{0.45\textwidth}
\includegraphics[width=\textwidth]{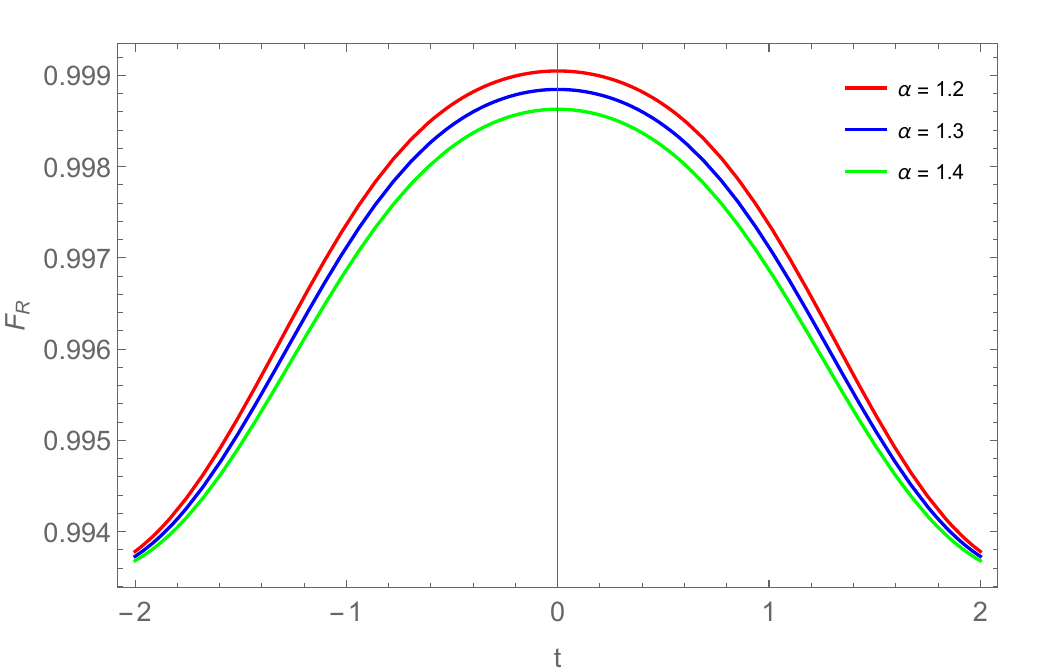}
\endminipage\hfill
\minipage{0.45\textwidth}
\includegraphics[width=\textwidth]{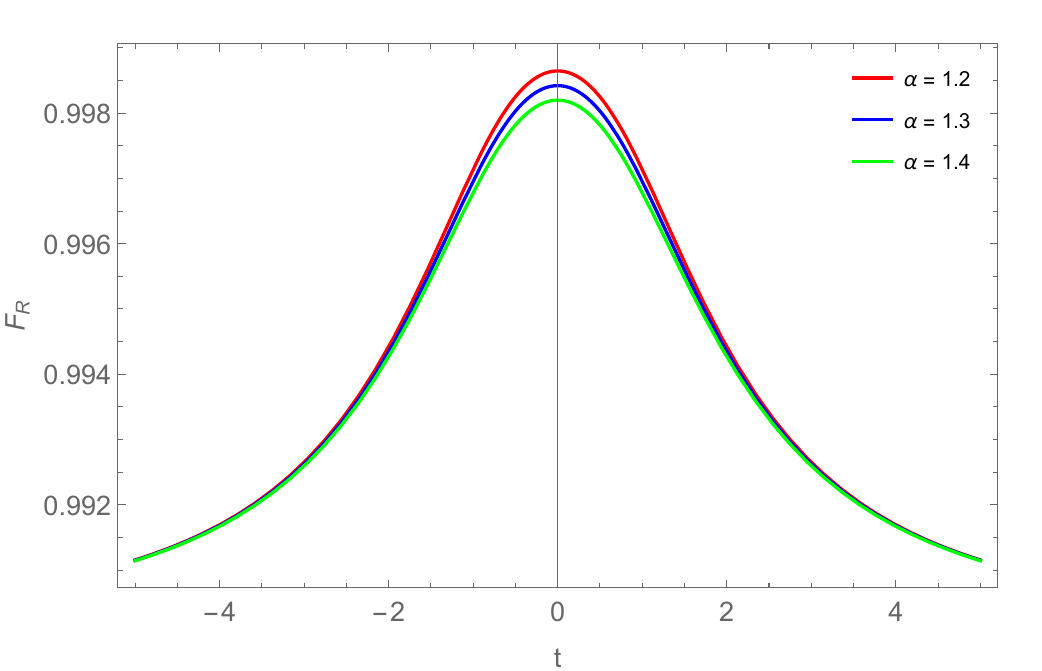}
\endminipage\hfill
\caption{Variations of $F_{R}=\frac{dF}{dR}$ vs. $t$ for Starobinsky model (left panel) and Exponential model (right panel) represented for different value of $\alpha$ with $\beta =0.9$}
\label{Fig.8}
\end{figure}

\section{Results and Conclusion}
 Two cosmological models of the Universe namely the Starobinsky $(n=1)$ \cite{Starobinsky07} and the exponential gravity model \cite{Cognola08} have been presented in $F(R)$ theory of gravity. The $F(R)$ theory of gravity is derived from an action where the usual Ricci scalar is replaced by a minimally coupled function in $R$. Two well-recognized forms of $F(R)$ function have been considered with a bouncing scale factor. Both Model I and Model II show the bouncing behaviour at $t=0$. Moreover, the values of the model parameters and the scale factor parameter are chosen so that the effective energy density of the models lies in the favourable profile. On the other hand, the effective pressure lies in the negative profile throughout the evolution of the universe. Furthermore, for the chosen bouncing scale factor, the value of parameter $\beta=0.9$ makes $n>1/2$ in $(a_{0}t_{F}^{2}+1)^{n}$ with $\alpha=1.2, 1.3$ and $1.4$ value for which the Hubble radius diverges at the bounce point and falls monotonically on both sides of the bounce before asymptotically shrinking to zero, indicating an accelerating late-time Universe \cite{Odintsov20a}. Also, such a behaviour of the scale factor is required for the compatibility of the given $F(R)$ theory with Planck constraints and generates the required perturbation modes near the bounce. From the behaviour of the cosmographic parameters, it has been noticed that the deceleration, jerk, snap and lerk parameter has a singularity at the bounce epoch. While the deceleration parameter is in the lower part of the zero line throughout the evolution confirms the accelerating nature of the Universe.  The EoS parameter curve crosses two times the phantom-divide line in both models, to support the bouncing behaviour. Further, the accelerated expansion of the models validated through the behaviour of EoS and deceleration parameters. We wish to mention here that, the presence of a finite non-zero value of $R_{0}$ throughout the bouncing epoch removes the singularity in the EoS parameter.\\

Further, the behaviour of EoS parameter also determined by the scale parameter of the scale factor. The violation of NEC and SEC in both the models are shown. These violations are inevitable in the context of modified theory of gravity and the bouncing scale factor. To note, the phantom phase might develop in the model with a positive Hubble parameter slope due to the violation of null energy requirements. Moreover stability of the model has been detected from the behaviour of the $F_{R}$ with cosmic time, both the models shows stable behaviour throughout the evolution. In conclusion, we comment that these two models may give some more insight in resolving the initial singularity issue of early Universe.\\

A.S. Agrawal: Conceptualization, Methodology, Calculations, Drafting, Plottings, S. Mishra: Conceptualization, Methodology, Calculations, S.K. Tripathy: Conceptualization, Methodology, Review, Editing, B. Mishra: Conceptualization, Methodology, Review, Editing.

The authors declare that they have no known competing financial interests or personal relationships that could have appeared to influence the work reported in this paper.

\section*{Acknowledgement} ASA acknowledges the financial support provided by University Grants Commission (UGC) through Senior Research Fellowship (File No. 16-9 (June 2017)/2018 (NET/CSIR)), to carry out the research work. BM and SKT thank IUCAA, Pune, India for providing support through visiting associateship program. The authors are thankful to the honorable referee for the constructive comments and suggestions for the improvement of the paper.\\

\end{document}